\documentclass[twocolumn,prb,showpacs,multicol,amsmath,amssymb]{revtex4}
\usepackage[dvips]{graphicx}
\usepackage{graphicx}
\usepackage{dcolumn}
\usepackage{bm}
\usepackage{graphics}
\usepackage{epsfig,color}
\usepackage[normalem]{ulem} 
\usepackage{soul}

\newcommand{\be}{\begin{equation}}
\newcommand{\ee}{\end{equation}}
\newcommand{\bea}{\begin{eqnarray}}
\newcommand{\eea}{\end{eqnarray}}

\newcommand{\bS}{\bf S}

\begin{document}
\title{Non-Markovian dynamics in the extended cluster spin-1/2 XX chain}
\author{M. Mahmoudi$^{1}$}
\author{M. R. Soltani$^{2}$}
\author{T. Mohammad Ali Zadeh$^{1}$}
\author{S. Mahdavifar$^{1}$}
\affiliation{$^{1}$ Department of Physics, University of Guilan, 41335-1914, Rasht, Iran}
\affiliation{$^{2}$ Department of Physics, Share-rey Branch, Islamic Azad University, Tehran, Iran}

\date{\today}
\begin{abstract}

 We study the dynamics of entanglement in the extended cluster spin-1/2 XX chain, equivalent
to a 1D spin-1/2 XX model with three-spin interaction (TSI). Selecting the nearest neighbor pair spins as an open quantum system, the rest of the chain plays the role of environment. The two-spin Heisenberg and the TSI interaction are responsible for coupling between system and environment. We show the existence of a critical value in the TSI, where the dynamics of concurrence changes from Markovian to the non-Markovian. In the region with the non-Markovian dynamics, entanglement sudden death in the system is observed. By focusing on the nearest neighbor pair spins of the environment, we have showed that the dynamics of entanglement in the environment is sensitive to the Markovian and non-Markovian regions.

 \end{abstract}
\maketitle

\section{Introduction}\label{sec1}

Recently, the growing interest in the dynamics of open quantum systems was amplified by entanglement sudden death (ESD) phenomenon\cite{Yu06, Eberly07, Papp09, Yu09, Barreiro10, Aolita15}, surprising discovery in the dynamical behavior of open systems uncovered  theoretically and experimentally. In principle, the interaction between system and environment degrades the quantum correlation in the entangled system and provokes the disappearance of entanglement in a finite time which is in sharp contrast to half-life law. To make this feature better understood,  the dynamics of entanglement for interacting two-level systems under the action of local stochastic environments was considered. Zyczkowski et al has inferred that revivals of entanglement is feasible\cite{Zyczkowski01}. It is worth noticing fact that the dynamics of single-particle  is asymptotical which it is not similar to ESD character\cite{Yu04, Santos06}.

Using the tools of open quantum systems theory, the dynamics of the reduced system state is typically categorized  as Markovian and non-Markovian\cite{Breuer07-1}. The process by which information flows only from the system to the environment is referred to as Markovian which was significantly successful in the frontier of quantum optics. What is more, in  Markovian dynamics, the concurrence decays exponentially and asymptotically\cite{Santos06}. However, when information flows from the system to the environment and vice versa, the dynamics is non-Markovian. Soft or condensed matter systems is the best platform for non-Markovian process describing strong interaction between system and environment. Therefore, the studies on the dynamics of open systems, which initially were confined to Markovian approximations, have recently been driven to non-Markovian environments\cite{Bellomo07, Piilo08, Breuer09,  Appolaro11, Madsen11, Liu11, Franco12,  Huelga12, Barnes12, Franco13, Xu13, Addis14, Orieux15}. In these works, researches are conducted on  particles interacting in different environments.
Individual, independent, common and yet combinations of both situations may be considered as an environment in the quantum optics. The most noticeable result observed for systems in contact with non-Markovian environments is revivals of entanglement \cite{Bellomo07, Franco12, Franco13, Xu13}. A quantum jump method for treating the dynamics of open systems that interact with non-Markovian environment is presented\cite{Bellomo07}. Given what mentioned above, it is necessary to also introduce a general measure for the degree of non-Markovian behavior in open systems\cite{Breuer09}.

In an intriguing  research, the dynamics of a qubit coupled to a spin chain environment is studied \cite{Appolaro11}. The spin chain environment is described by an XY model in a transverse magnetic field. In the parameter space of the system, there is a specific point where the qubit dynamics is Markovian. Separated into two regions, this point triggers two totally different dynamical behaviors. In addition, a system of dimension $N$  are separated into two parts. In other words, there are a single qubit  and the other part as an environment strongly coupled to qubit\cite{Marko11}. It is found that the contribution due to energy density is responsible for non-Markovian effects even in the limit of an infinite environment. The role of environment quantum correlations on the evolution of a spin chain is also studied\cite{Lorenzo11}. It is argued that the presence of entanglement in the state of the environmental system expedites the non-Markovian character of the chain's dynamics. Recently, the dynamics of a central qubit coupled to a quantum Ising ring in the transverse field is investigated\cite{Haikka12}. According to the recent study, it has been found that environmental criticality has strong impact on the information flux between system and environment. As an indicator of criticality, non-Markovianity plays key role in this model.

The main purpose of this paper is to investigate the effect of the three-spin interaction (TSI) on the dynamics of quantum correlations between the nearest neighbor pair spins in a 1D spin model. The motivation behind this study is the recent progress in the field of quantum magnets. It is known that a wide variety of novel spin-1/2 Hamiltonians can be generated in the different configurations of an optical lattice\cite{Kuklov03, Duan03}. Take a triangular configuration as an example, which can propose an effective three-spin interaction (TSI) \cite{Pachos04}. The one-dimensional spin-1/2 model with added TSI can precisely describe the dynamics of two species of bosons trapped in an optical lattice with a triangular-ladder configuration\cite{Dcruz05}. It has shown that the TSI has positive effects on the computational power of one-way quantum computation\cite{Tame06}. Therefore, we consider a spin-1/2 Heisenberg XX chain model with TSI. By selecting a nearest neighbor pair spins as an open quantum system, the others play as an environment (Fig.1). In order to diagonalize the spin chain model, we apply fermionization technique in the thermodynamic limit.  According to results, the dynamics of the entangled system is Markovian in the absence of the TSI. However, there is furthermore subtle point we must consider, In the presence of the TSI, Markovian dynamics remains unchanged up to a critical value ($\alpha_c$) where non-Markovian dynamics shows up. We explicitly show that entanglement sudden death phenomenon happens in the region of non-Markovian dynamics. To cast more clarification on the offered points,  we investigate the propagation of entanglement in the environment. The most persuasive point is that the environment has been affected by non-Markovian dynamics as expected intuitively.

The paper is organized as follows. In the next section we introduce the model and find an analytical form for the entanglement and the QD. In section \ref{sec3}, analytical results will be presented. Finally, we conclude and summarize our results in section \ref{sec4}.

\begin{figure}[t]
\centerline{\psfig{file=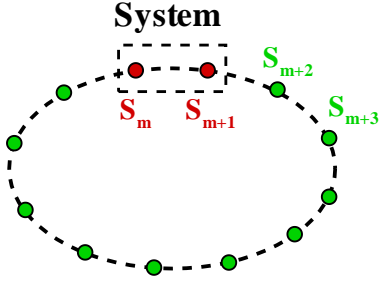,width=3.5in}}
\caption{(color online). The schematic diagram of the  spin chain model.}
\label{Fig-0}
\end{figure}

\section{The model and the concurrence}\label{sec2}

We propose the simplest form of the generalized XY models so called the extended XY chain model spin-1/2 with three spin interaction, fully characterized  by zig-zag chain as\cite{Titvinidze03}
\begin{eqnarray}
{\cal H} &=& J \sum_{n=1}^{N} (S^{x}_{n}
S^{x}_{n+1}+S^{y}_{n} S^{y}_{n+1})\nonumber \\
&-& J'  \sum_{n=1}^{N}
(S_{n}^{x} S_{n+1}^{z} S_{n+2}^{x}+S_{n}^{y} S_{n+1}^{z} S_{n+2}^{y}),   \label{Hamiltonian}
\end{eqnarray}
which can be rewritten as
\begin{eqnarray}
{\cal H} &=& {\cal H}_S+{\cal H}_E+{\cal H}_{SE},\nonumber \\
{\cal H}_S&=& J (S^{x}_{m} S^{x}_{m+1}+S^{y}_{m} S^{y}_{m+1}),\nonumber \\
{\cal H}_E&=& -J' \sum_{n\neq m-2,m-1,m,m+1}(S_{n}^{x} S_{n+1}^{z} S_{n+2}^{x}\nonumber \\
&+&S_{n}^{y} S_{n+1}^{z} S_{n+2}^{y})+J \sum_{n=m+1}^{N} (S^{x}_{n}
S^{x}_{n+1}+S^{y}_{n} S^{y}_{n+1}),\nonumber \\
{\cal H}_{SE}&=&J (S^{x}_{m-1} S^{x}_{m}+S^{y}_{m-1} S^{y}_{m}+S^{x}_{m+1} S^{x}_{m+2}\nonumber \\
&+&S^{y}_{m+1} S^{y}_{m+2})\nonumber \\
&-&J' \sum_{n=m-2}^{m+1}(S_{n}^{x} S_{n+1}^{z} S_{n+2}^{x}+S_{n}^{y} S_{n+1}^{z} S_{n+2}^{y}),
\label{Hamiltonian}
\end{eqnarray}
  where the sum over n satisfying the periodic boundary conditions goes from $1$ to $N$,  $S_{n}$ is used to denote the spin-1/2 operator on the $n$-th site. The transverse exchange coupling between the spins on the nearest-neighbor sites is denoted by $J$, while  transverse exchange between the spins on the next-nearest neighbor sites is introduced by $J^{'}$, which depends on "$z$" orientation of the spin surrounded by the next-nearest-neighbors. It should be noted that, one can easily reconstruct the features of the system for $J < 0$ using the transformation $S^{x,y}_{n}\rightarrow (-1)^n S^{x,y}_{n}$, $S^{z}_{n}\rightarrow S^{z}_{n}$. On the other hand, since the sign of the three spin coupling term is changed by the time reversal transformation $S^{y}_{n}\rightarrow S^{y}_{n}$, $S^{x,z}_{n}\rightarrow - S^{x,z}_{n}$, the properties of the system for $J' < 0$ can be easily obtained from the case $J' > 0$.

This model is exactly solveable\cite{Titvinidze03}. Using the Jordan-Wigner transformation
\begin{eqnarray}
S_{n}^{z}&=&a_{n}^{\dagger}a_{n}-\frac{1}{2}, \\ \nonumber
S_{n}^{+}&=&a_{n}^{\dagger} \exp(i\pi\sum_{l<n}a_{l}^{\dagger}a_{l}),\\ \nonumber
S_{n}^{-}&=&a_{n} \exp(-i\pi\sum_{l<n}a_{l}^{\dagger}a_{l}),
\label{Hamiltonian}
\end{eqnarray}
spins are  mapped  onto a one-dimensional noninteracting spinless fermions with creation and annihilation operator
\begin{eqnarray}
H_{f}&=&\frac{J}{2} \sum_{n} (a^{\dag}_{n}a_{n+1}+ h.c.)\nonumber \\
&-&\frac{J'}{4} \sum_{n} (a^{\dag}_{n}a_{n+2}+ h.c.).
\end{eqnarray}
By performing a Fourier transformation into the momentum space as $a_{n} = \frac{1}{\sqrt{N}} \sum ^{N} _{n=1} e^{-ikn} a_{k}$, the diagonalized Hamiltonian is given by
\begin{eqnarray}
{\cal H}=\sum_{k}\varepsilon(k) a_{k}^{\dagger}a_{k}.
\label{Hamiltonian-d}
\end{eqnarray}
where $\varepsilon(k)$ is the dispersion relation and $\alpha=\frac{J^{'}}{J}$
\begin{eqnarray}
\varepsilon(k)=\cos(k)+\frac{\alpha}{2} \cos(2~k).
\end{eqnarray}
In what follows, we try to determine the dynamics of entanglement  in the zigzag chain. Specially, we concentrate on the pairwise entanglement between the nearest neighbor pair spins located at sites $m$ and $m+1$ in the chain system. For this purpose, we consider the initial state in which two sites ($m$ and $m+1$) in the chain system are maximally entangled. By doing so, the rest of the chain (the environment) is disentangled in this configuration as
\begin{eqnarray}
|\psi(t=0)\rangle&=&\frac{1}{\sqrt{2}}(|\uparrow\downarrow\rangle+e^{i\phi}|\downarrow\uparrow\rangle)_S \\ \nonumber
&\otimes&(|\downarrow\downarrow\downarrow...\downarrow\rangle)_E,
\label{initial-state-0}
\end{eqnarray}
which is equivalent to
\begin{eqnarray}
|\psi(t=0)\rangle=\frac{1}{\sqrt{2}} (a_{m}^{\dagger}|0\rangle+a_{m+1}^{\dagger}e^{i\phi}|0\rangle),
\label{initial-state}
\end{eqnarray}
in the fermion language. $|0\rangle$ denotes the vacuum state and $\phi$ is a phase factor. Using the time evolution operator, $U(t)=e^{\frac{-it}{\hbar}\sum_{k}\varepsilon(k) a_{k}^{\dagger}a_{k}}$, the physical state of the system at time $t$ ($\hbar=1$ is considered) is obtained as
\begin{eqnarray}
|\psi(t)\rangle=\frac{1}{\sqrt{2 N}} \sum_{k} f(k)a_{k}^{\dagger}|0\rangle,
\label{physical-state}
\end{eqnarray}
where $f(k)=e^{i(km-\varepsilon(k) t)}(1+e^{i(k+\phi)})$. We give a detailed analysis to facilitate the understanding of entanglement regarding the concurrence. As mentioned in most contexts, the concurrence is the indicator of the entanglement measure for any bipartite system . Therefore, our study is focused on the concurrence of two spins at sites $i$ and $j$. The concurrence can be  illustrated by the corresponding reduced density matrix $\rho_{ij}$, which in
the standard basis is expressed as
\begin{eqnarray}
\rho_{ij}= \left(
             \begin{array}{cccc}
               <P_{i}^{\uparrow}P_{j}^{\uparrow}> & <P_{i}^{\uparrow}{\bS}_{j}^{-}> & <{\bS}_{i}^{-}P_{j}^{\uparrow}> & <{\bS}_{i}^{-}{\bS}_{j}^{-}> \\
               <P_{i}^{\uparrow}{\bS}_{j}^{+}> & <P_{i}^{\uparrow}P_{j}^{\downarrow}> & <{\bS}_{i}^{-}{\bS}_{j}^{+}> & <{\bS}_{i}^{-}P_{j}^{\downarrow}> \\
               <{\bS}_{i}^{+}P_{j}^{\uparrow}> & <{\bS}_{i}^{+}{\bS}_{j}^{-}> & <P_{i}^{\downarrow}P_{j}^{\uparrow}> & <P_{i}^{\downarrow}{\bS}_{j}^{-}> \\
               <{\bS}_{i}^{+}{\bS}_{j}^{+}> & <{\bS}_{i}^{+}P_{j}^{\downarrow}> & <P_{i}^{\downarrow}{\bS}_{j}^{+}> & <P_{i}^{\downarrow}P_{j}^{\downarrow}> \\
             \end{array}\nonumber
           \right). \label{density matrix1}
\end{eqnarray}

where $P^{\uparrow}=\frac{1}{2}+S^{z}, P^{\downarrow}=\frac{1}{2}-S^{z}$. The brackets denote the expectation  values at time $t$ and  $S^{\pm}= S^{x}\pm i S^{y}$. The concurrence between two spins is given via  $C_{i,j}=max(0,\lambda_{1}-  \lambda_{2} -\lambda_{3} -\lambda_{4}$) where  $\lambda_{i}$ is the square root of the eigenvalue of $R=\rho_{i,j}  \tilde{\rho}_{i,j}$ and  $\tilde{\rho}_{i,j}=(\sigma_{i}^{y}\otimes\sigma_{j}^{y})~\rho^{\ast}(\sigma_{i}^{y}\otimes\sigma_{j}^{y})$. By using the Jordan-Wigner transformation, the reduced density matrix for NN spins will be given by

\[
\rho_{m,m+1} =
\left( {\begin{array}{cccc}
  X^{+}  & 0 &  0 & 0\\
    0 &  Y^{+}  & Z^{*} & 0\\
 0 & Z & Y^{-}  & 0\\
 0 & 0 & 0 & X^{-}
 \end{array} } \right),
\]
where $X^{+}=\langle n_{m} n_{m+1}\rangle~(n_m=a_{m}^{\dagger}a_{m})$,  $Y^{+}=\langle n_m(1-n_{m+1})\rangle$, $Y^{-}=\langle n_{m+1}(1-n_{m})\rangle$, $Z=\langle a_{m}^{\dagger}a_{m+1} \rangle$ and $X^{-}=\langle 1-n_m- n_{m+1}+n_m n_{m+1}\rangle$. Thus the concurrence is transformed into
\begin{eqnarray}
C_{m,m+1}=max\{0,2 (|Z|-\sqrt{X^{+} X^{-}})\},
\label{concurrence}
\end{eqnarray}
where
\begin{eqnarray}
Z=\frac{1}{8 \pi^{2}} \int_{-\pi}^{\pi} \int_{-\pi}^{\pi} f^{\star}(k)~f(k')e^{i(k-k')m-k'} dk~dk',
\end{eqnarray}
\begin{eqnarray}
\langle n_{m}\rangle=\frac{1}{8 \pi^{2}} \int_{-\pi}^{\pi} \int_{-\pi}^{\pi} f^{\star}(k)~f(k')e^{i(k-k')m} dk~dk',
\end{eqnarray}
\begin{eqnarray}
X^{+}=\langle n_{m}\rangle~\langle n_{m+1}\rangle-Z Z^{\star}.
\end{eqnarray}

\begin{figure}[t]
\centerline{\psfig{file=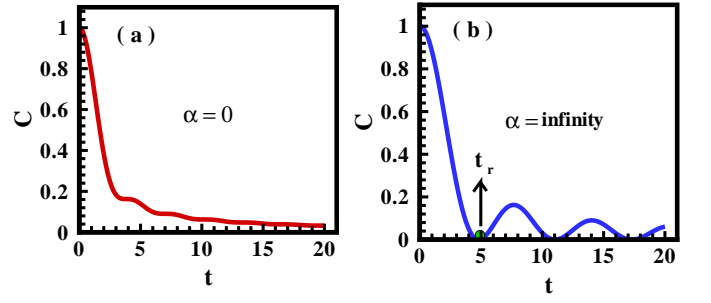,width=3.5in}}
\caption{(color online). The time behavior of concurrence in the system, $C_{m, m+1}$, for (a)  Heisenberg  interaction $J'=0$ $(\alpha=0)$ and (b)  TSI interactions $J=0.0$ $(\alpha=infinity)$.}
\label{Fig-1}
\end{figure}

\section{measure of the entanglement}\label{sec3}

In  the following, we evaluate the concurrence between NN pair spins $m$ and $m+1$ for different values of $\alpha$, and describe the behavior of the concurrence with respect to the time. We consider the case, in the chain, where NN spins on sites $m$ and $m+1$ are initially prepared in a maximally entangled Bell state, as defined in Eq.~(\ref{initial-state-0}) ($\phi=0$ is considered), and look at the concurrence between them with respect to the time.

 The graph in Fig.~\ref{Fig-1} provides information about the dynamical behavior of the concurrence in the absence of the TSI (a) and  the Heisenberg interaction (b).  In general, the time evolution induced by the Hamiltonian is totally different in these cases. There is a downward trend in the concurrence behavior by varying degrees.

As regards the first, in Fig.~\ref{Fig-1} (a), the concurrence declines moderately and decays at longer times as $t^{-1}$. To put it more simply, in the absence of the TSI, the system is connected to its environment with the two-point Heisenberg interaction. As a matter of fact, the two-point Heisenberg interaction between system and  environment  (the rest of the chain) causes flipping  spins in the chain, resulting in a flow of the entanglement into the environment which is known as an indication of the Markovian dynamics. In the contrary, in Fig.~\ref{Fig-1} (b), when the two-point Heisenberg interaction does not exist, the entanglement drops sharply by passing time and disappears in the finite time compared to Fig.~\ref{Fig-1} (a) (in the first time at $t_{r}=4.80\pm0.02$). The most striking fact is that the entanglement sudden death called as ESD phenomenon emerges in the presence of the TSI. It is noticeable that as soon as the time passes from $t_{r}$, the concurrence  regains and peaks, then decreases and reaches to zero for the second time. The mentioned behavior is repeated periodically in time which is also known as the revival of the entanglement. Such a revival is due to the special action of the environment which is created through the TSI. In principle, the feedback of the quantum correlations from the environment into the system, characterizing non-Markovian dynamics, enhances the appearance of ESD regions.

\begin{figure}[t]
\centerline{\psfig{file=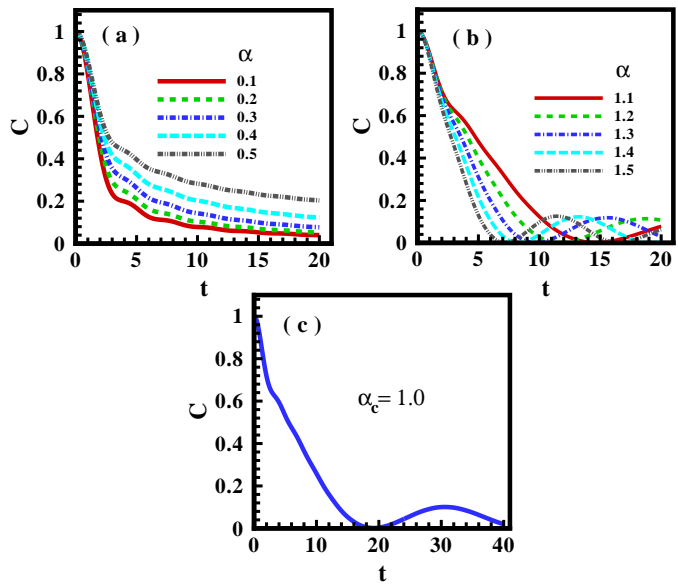,width=3.5in}}
\caption{(color online). The time behavior of  concurrence in the system, $C_{m, m+1}$,  for different values of the TSI interactions (a) $\alpha<\alpha_c$, (b) $\alpha>\alpha_c$ and (c) $\alpha=\alpha_c$.
}
\label{Fig-2}
\end{figure}

Since the dynamics of the system changes from Markovian to the non-Markovian, it is completely natural to search for a critical value($\alpha_c$) in the TSI . To find it, we have calculated the concurrence between NN pair spins $m$ and $m+1$ as a function of the time and $\alpha$. The graphs, in Fig.~\ref{Fig-2}, compare figures for different values of $\alpha$.
As it can be seen from Fig.~\ref{Fig-2} (a), for the values less than the critical value $\alpha<\alpha_{c}=1.0$, the concurrence decays asymptotically with respect to the time. It is interesting to note that for $\alpha>\alpha_{c}$, the pattern is not similar to Fig.~\ref{Fig-2} (a), the strength of exchange coupling $J^{'}$ may contribute to a flow of the entanglement into the environment (see Fig.~\ref{Fig-2}(b)). As a result, the dynamics of the system is non-Markovian in the region $\alpha>\alpha_{c}=1.0$. For simplicity, $J=1$ is considered. We have also plotted the time behavior of the concurrence exactly at the critical TSI $\alpha_{c}=1.0$ in Fig.~\ref{Fig-2} (c). By comparing results presented in Fig.~\ref{Fig-2} (b) and (c), it can be inferred that by increasing in the TSI via critical value, $t_{r}$ decreases.

\begin{figure}[t]
\centerline{\psfig{file=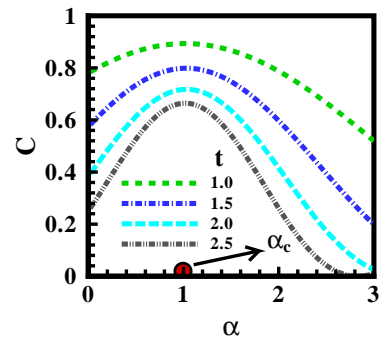,width=2.0in}}
\caption{(color online). The behavior of the concurrence in the system, $C_{m, m+1}$, as a function of the $\alpha$ in different times $t<t_r$. The concurrence is maximized exactly at the critical value of the TSI $\alpha_c$.}
\label{Fig-3}
\end{figure}

To confirm the existence of the aforementioned phase transition, we have also investigated the static behavior of the concurrence with respect to the TSI ($\alpha$). Fig.~\ref{Fig-3} gives results for different values of time less than $t_{r}$. It is clearly seen that static behavior of the concurrence shows a peak exactly at the critical TSI ($\alpha=\alpha_c=1$). As a matter of fact, at the critical point ($\alpha=\alpha_c$), the rate of concurrence decline  at short times is minimum, which results from this fact that entanglement has a flow from the environment into the system in the non-Markovian dynamics.

\begin{figure}[t]
\centerline{\psfig{file=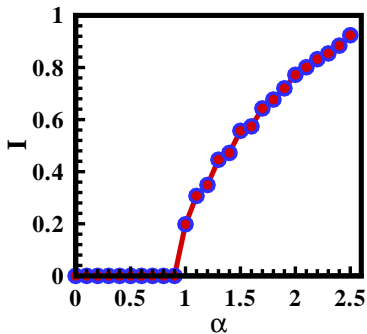,width=2.0in}}
\caption{(color online). The witness of non-Markovianity as a function of the TSI, clearly seen the non-Markovian dynamics in the region $\alpha>\alpha_c$. }
\label{Fig-4}
\end{figure}

In recent years, wide interest is devoted to the quantification of the degree of non-Markovianity of a dynamical evolution. In particular, people have focused on the identification of appropriate tools for the
characterization of the many facets of non-Markovianity. Here, we utilize one of the special measures\cite{Rivas10} to determine the features of the dynamics under study here. Based on this method, by computing the amount of entanglement between
the two parties of the system at different instants of times within a selected interval $[t_0, t_{max}]$, one can detect the non-markovianity behavior. The witness of non-Markovianity is defined as

\begin{eqnarray}
I=\int_{t_{0}}^{t_{max}}|\frac{dC}{dt}| dt - \Delta C,
\end{eqnarray}
where $\Delta C=C(t_0)-C(t_{max})$. In the Markovian dynamics, the  first derivative of the concurrence is negative and thus witness will be zero. Fig.~\ref{Fig-4} depicts the witness of non-Markovianity as a function of the TSI. Here we considered $t_0=0$ and $t_{max}=40$. As shown, in the region $\alpha<\alpha_c$, the witness is zero as an indication of the Markovian evolution. It is clear from data that  by increasing TSI from  $0$ to $\alpha_c$, the witness kindles and increases smoothly.

\begin{figure}[t]
\centerline{\psfig{file=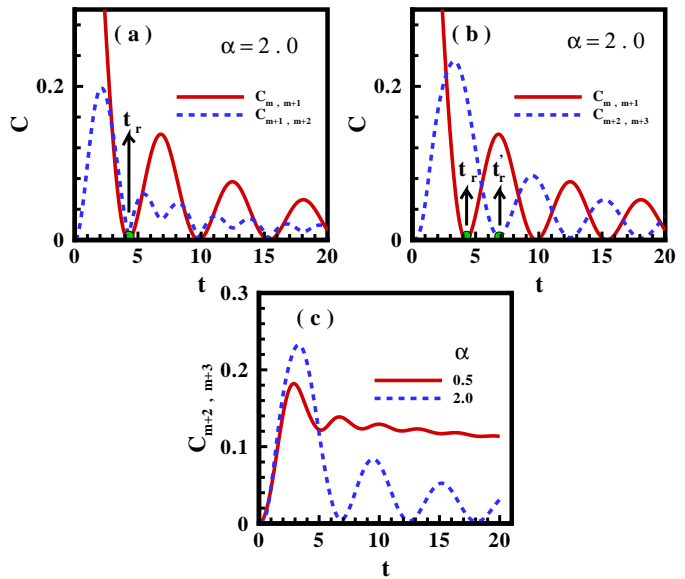,width=3.5in}}
\caption{(color online). The comparison between (a) $C_{m, m+1}$ and $C_{m+1, m+2}$ versus t, (b) The comparison between $C_{m, m+1}$ and $C_{m+2, m+3}$ versus $t$ (c) The results of  concurrence between pair spins of the environment adjacent to the system ($C_{m+2, m+3}$) for different TSI, $\alpha=0.5$ $(\alpha<\alpha_c)$ and $\alpha=2.0$ $(\alpha>\alpha_c)$. }
\label{Fig-5}
\end{figure}

 To find a deeper insight into the nature of  non-Markovian dynamics, we concentrated on the concurrence between the edge spins of the system and the environment ($C_{m+1, m+2}$), as well as the concurrence between pair spins of the environment adjacent to the system  ($C_{m+2, m+3}$) (see Fig.~\ref{Fig-5}). Undoubtedly, we expect that the time evolution dictated by the Hamiltonian, amounts to a simultaneous spin flip between NN spins in the environment; consequently, entanglement propagates into the environment. In Fig.~\ref{Fig-5}, we plot (a) $C_{m, m+1}$ and $C_{m+1, m+2}$ versus time, (b) $C_{m, m+1}$ and $C_{m+2, m+3}$ versus time for the amount $\alpha=2.0>\alpha_c$. It can be seen from Fig.~\ref{Fig-5} (a), at the initial time (t=0), there is not any quantum correlation between the edge spins of the system and the environment, that is,  $C_{m+1, m+2}=0$, which is in complete agreement with the meaning of the initial state (disentangled state) Eq.~(\ref{initial-state-0}). With the passage of time, the edge spins of the system will be entangled. As a result concurrence ($C_{m, m+1}$) reaches to a maximum value and then dips to zero for the first time ($t_r$). There is no cast of doubt on very fact that the concurrence between spin pairs of the system is the same as the concurrence between the edge spins at the time ($t_{r}$) with the $C=0$. In fact, TSI interaction between system and the environment will cause no concurrence between pair spins in the system and between edge spins of the system and the environment. The graph.~\ref{Fig-5} (b) compares figures for the concurrence of the system ($C_{m, m+1}$), together with the concurrence between pair spins of the environment adjacent to the system ($C_{m+2, m+3}$). It holds great significance to notice that the ESD phenomenon is also clearly seen in the environment when the dynamics is non-Markovian ($\alpha=2.0$). On the other hand, the time, where the concurrence between pair spins of the environment disappears, ($t^{'}_{r}$) is completely different from the system ($t_{r}$). Obviously, the entanglement of the system is maximized when the concurrence between pair spins of the environment adjacent to the system disappears. Finally, the graph.~\ref{Fig-5} (c) gives figures for the concurrence between pair spins of the environment adjacent to the system  versus time for two different values TSI, $\alpha=0.5<\alpha_c$ and $\alpha=2.0>\alpha_c$. In a nutshell, there is no evidence of ESD in the system with Markovian dynamics.

\section{Conclusion}\label{sec4}

In  this  paper, we  have  studied the  dynamics of entanglement between the two-nearest neighbor spins in the 1D spin-1/2 with TSI.  We have implemented the fermionization technique to find analytical results. We selected the nearest neighbor pair spins as an open quantum system. It is obvious that the rest of the chain plays the role of the environment. The desired quantum open system can be coupled to the environment via both two-spin Heisenberg and TSI interaction.

We showed that the dynamics of concurrence of the system is Markovian when TSI is absent. By adding TSI to the Hamiltonian, we found  a critical value in the TSI ($\alpha_c$), where the dynamics of concurrence varies from Markovian to non-Markovian. We also observed that entanglement sudden death of  the system emerged in the region with non-Markovian dynamics. On the other hand, we utilized a measure to elucidate the degree of non-Markovianity and determined the features of the dynamics under study.

In addition, by focusing on the propagated entanglement in the environment, we showed that the dynamics of entanglement in the environment is sensitive to the Markovian and non-Markovian regions. Based on results, the concurrence between the the edge spins of the system and the environment disappears exactly at the same time which is similar to the case has been reported about the system. The most telling conclusion to be drawn is that ESD phenomenon was observed in the environment for non-Markovian dynamics. On the other hand, the time, where the concurrence between pair spins of the environment disappears, is absolutely different from the case in the system. At this point, one can see a peak in the system graph which is a manifestation of maximized entanglement.

\section{acknowledgments}


\vspace{0.3cm}


\end{document}